\documentclass[oneside,twocolumn]{article}

\usepackage{geometry}
\usepackage{fullpage}

\usepackage[utf8]{inputenc}
\usepackage{graphicx}
\usepackage{amsmath,mathtools}
\usepackage{amssymb}
\usepackage{amscd}
\usepackage{bm}
\usepackage{url}
\usepackage[super,sort&compress,comma]{natbib}
\usepackage{upgreek}
\usepackage{hyperref}
\usepackage{pdflscape}
\usepackage[font=small,labelfont=bf,labelsep=period]{caption}
\usepackage[version=3]{mhchem}
\usepackage{booktabs}
\usepackage{multirow}
\usepackage{threeparttable}
\usepackage{sectsty}
\usepackage{setspace}
\usepackage{authblk}
\usepackage{fancyhdr}
\usepackage{enumitem}
\usepackage{xcolor}

\usepackage[T1]{fontenc}
\usepackage{textcomp}
\usepackage[euler]{textgreek}

\makeatletter
  \def\@seccntformat#1{\csname the#1\endcsname.\ \ }
\makeatother

\sectionfont{\large}
\subsectionfont{\normalsize}

\makeatletter
 \def\@biblabel#1{#1.}
 \makeatother
\newcommand*{\rsub}[1]{\ensuremath{_{\mathrm{#1}}}}
\newcommand*{\rsup}[1]{\ensuremath{^{\mathrm{#1}}}}

\newcommand*{\degc}{\ensuremath{^\circ\mathrm{C}}}
\newcommand*{\degree}{\ensuremath{^\circ}}
\newcommand*{\pers}{\ensuremath{\mathrm{s}^{-1}}}
\newcommand*{\tmu}{\textmugreek}
\newcommand*{\us}{{\textmugreek}s}

\newcommand*{\htwo}{\ce{^{2}H}}
\newcommand*{\ctt}{\ce{^{13}C}}

\newcommand*{\mn}{\ce{Mn^{2+}}}
\newcommand*{\ca}{\ce{Ca^{2+}}}

\newcommand*{\mM}{m\textsc{m}}

\makeatletter
\def\@maketitle{%
  \newpage
  \begingroup
    \let \footnote \thanks
    \hrule height \z@        
    {\LARGE \bfseries \@title \par}%
    \vskip 6mm
    {\large
     \@author
    }%
  \par\endgroup
  \vskip 5mm
}
\makeatother

\title{Mobility of core water in \textit{Bacillus subtilis} spores by \htwo\ 
       NMR}

\author[1]{Shuji Kaieda}
\author[2]{Barbara Setlow}
\author[2]{Peter Setlow}
\author[1]{Bertil Halle}

\affil[1]{Department of Biophysical Chemistry, Lund University, 
       Lund, Sweden}
\affil[2]{Department of Molecular, Microbial and Structural Biology, 
          University of Connecticut Health Center, 
          Farmington, Connecticut, USA}

\date{}

\begin{document}

%
%
\twocolumn[
\begin{@twocolumnfalse}

\maketitle\thispagestyle{plain}

%
%
\noindent Bacterial spores in a metabolically dormant state can survive long 
periods without nutrients under extreme environmental conditions. 
The molecular basis of spore dormancy is not well understood, but the 
distribution and physical state of water within the spore is thought to play an 
important role. 
Two scenarios have been proposed for the spore's core region, containing the DNA 
and most enzymes. 
In the gel scenario, the core is a structured macromolecular framework permeated 
by mobile water. 
In the glass scenario, the entire core, including the water, is an amorphous 
solid and the quenched molecular diffusion accounts for the spore's dormancy and 
thermal stability. 
Here, we use \htwo\ magnetic relaxation dispersion to selectively monitor water 
mobility in the core of \textit{Bacillus subtilis} spores in the presence and 
absence of core \mn\ ions. 
We also report and analyze the solid-state \htwo\ NMR spectrum from these 
spores. 
Our NMR data clearly support the gel scenario with highly mobile core water 
($\sim 25$~ps average rotational correlation time). 
Furthermore, we find that the large depot of manganese in the core is nearly 
anhydrous, with merely 1.7~\% on average of the maximum sixfold water 
coordination. 
\vspace{0.5cm}
\end{@twocolumnfalse}
]

%
%
\section{\label{sec:intro}Introduction}

The common Gram-positive bacterium \textit{Bacillus subtilis} can survive a 
variety of adverse environmental conditions by transforming into a 
metabolically dormant endospore.\cite{Gerhardt1989,Setlow2006} 
The molecular basis of spore dormancy is not well understood, but may provide a 
key to controlling food spoilage and food-borne disease caused by bacterial 
spores of certain species. 
Specifically, mechanistic insights may lead to more efficient methods for 
inactivating spores.\cite{Mafart2010,Eijlander2011}

In the spore, the bacterial genome and most of the essential enzymes are 
confined to a core region surrounded by an inner membrane (IM) with low 
permeability to water, ions, and small solutes.\cite{Setlow2006,Cowan2004,
Sunde2009b} 
The space outside the IM, known as the cortex, contains a cross-linked 
peptidoglycan matrix that is thought to maintain the lower water activity in the 
core.\cite{Popham2002}
The cortex is surrounded by the coat, another protective barrier composed of 
multiple layers of cross-linked proteins.\cite{Driks1999,Henriques2000} 

The amount and physical state of water in the different spore compartments, in 
particular the core, is thought to be crucial for spore dormancy and 
resistance.\cite{Gerhardt1989,Setlow2006} 
The core is less hydrated than the cortex and its water content is inversely 
related to the spore’s heat resistance.\cite{Nakashio1985,Beaman1986} 
It has been proposed that the entire core exists in an amorphous solid-like or 
glassy state.\cite{Gould1986,Sapru1993,Ablett1999} 
If so, the suppression of molecular diffusion could explain spore dormancy and 
the spore's heat resistance could be rationalized in kinetic terms.  
Reduced mobility has been demonstrated for most of the principal constituents of 
the core, such as ions,\cite{Carstensen1979,Johnstone1982} 
pyridine-2,6-dicarboxylic acid (dipicolinic acid, DPA),\cite{Ablett1999,
Leuschner2000} proteins,\cite{Cowan2003} and IM lipids,\cite{Cowan2004} but the 
mobility of core water has remained controversial.

Early on, it was observed that virtually all spore water exchanges with external 
water within minutes,\cite{Black1962,Marshall1970} implying that core water is 
in a liquid, rather than a glassy, state. 
This led to what may be termed the ``gel scenario'' of the core as a matrix of 
immobilized macromolecules (as well as DPA and chelated divalent metal ions) 
permeated by mobile or, at least, liquid water.\cite{Black1962} 
According to the competing ``glass scenario'', initially proposed as a 
speculation,\cite{Gould1986} the entire core, including the water, is 
immobilized. 
This view appeared to receive experimental support from differential scanning 
calorimetry results,\cite{Ablett1999} but alternative interpretations, not 
involving immobilized water, of these results have been 
suggested.\cite{Leuschner2003,Stecchini2006}. 
NMR can potentially probe water mobility more directly, but the early proton NMR 
studies\cite{Maeda1968,Bradbury1981} appear to have been confounded by the 
effects of paramagnetic Mn(II) ions. 

An electron paramagnetic resonance (EPR) spin-probe study used chemical and 
mechanical treatment to remove the outer parts of the spore and a paramagnetic 
broadening agent to separate the EPR signal from spin probes residing inside and 
outside the spore.\cite{deVries2006} 
In this way, the rotational correlation time of the spin probe could be inferred 
and related to the viscosity of spore water, which was found to exceed the bulk 
water viscosity by a factor of ten in the core and by a factor of five in the 
cortex.\cite{deVries2006} 
However, these results were obtained by an elaborate difference procedure, the 
quantitative validity of which hinges on several unproven assumptions. 
Furthermore, if the spin-probe interacts strongly with any of the spore 
components, its rotational motion might not reflect the mobility of the solvent.          

More recently, \htwo\ magnetic relaxation dispersion (MRD) measurements on 
\ce{D2O}-exchanged spores were used to directly monitor the rotational mobility 
of water molecules in the core and non-core regions.\cite{Sunde2009b} 
Importantly, this study demonstrated that water exchange across the IM is slow 
on the \htwo\ relaxation time scale ($\sim 10$~ms), thus allowing core and 
non-core water to be characterized individually in the same experiment. 
Consistent with the gel scenario, it was thus established that core water is 
highly mobile, with a rotational correlation time of 
$\sim 50$~ps.\cite{Sunde2009b}  
Although the core water content deduced from the MRD data\cite{Sunde2009b} was 
consistent with independent estimates,\cite{Nakashio1985} the existence of a 
small fraction of immobilized water could not be excluded since immobilized 
water does not contribute to the NMR signal detected in MRD experiments. 
Indeed, a subsequent \htwo\ solid-state NMR study recorded a broad feature in 
the \htwo\ NMR spectrum from partly dehydrated spores.\cite{Rice2011} 
This observation was taken to indicate a substantial amount of immobilized water 
in the spore,\cite{Rice2011} as predicted by the glass hypothesis.          

The principal aim of the present study is to resolve the long-standing issue of 
core water mobility by extending the \htwo\ MRD measurements and by performing 
\htwo\ solid-state NMR experiments on fully hydrated \textit{B.\ subtilis} 
spores. 
Specifically, by measuring the bi-exponential \htwo\ relaxation over a wider 
frequency range, we reveal the characteristic signature of Mn(II)-induced 
paramagnetic relaxation also for core water. 
Using also a spore preparation with $> 99$~\% of the paramagnetic \mn\ ions in 
the core substituted by diamagnetic \ca\ ions, we could quantify the small 
manganese-water contact in the core. 
These data also yielded an improved characterization of core water mobility, 
showing that the rotation of water molecules in the core is merely slowed down 
15-fold, on average, as compared to bulk water. 
We also examined spores that had been sporulated at different temperatures, 
finding a non-monotonic variation of the core water content. 
This finding, which deviates from the previously established reduction of core 
water content with increasing sporulation temperature,\cite{Beaman1986,
Melly2002} is tentatively taken to imply that additional factors influence the 
core water content. 
Finally, we demonstrate that the \htwo\ quadrupolar echo NMR spectrum from fully 
hydrated spores can be rationalized without invoking any immobilized spore water. 
Both our MRD and solid-state NMR results thus support the gel scenario for the 
core of \textit{B.\ subtilis} spores.

%
%
\section{\label{sec:methods}Materials and methods}

%
%
\subsection{\label{subsec:prep}Spore preparation and analysis}

{\small
The spore samples were prepared from \textit{B.\ subtilis} strain 
PS533.\cite{Setlow1996} 
Three samples were sporulated on standard 2 $\times$ SG medium agar plates 
at 23, 37, or 43~$\degc$ and a fourth sample was sporulated in an essentially 
Mn-free liquid 2 $\times$ SG medium at 
37~$\degc$.\cite{Paidhungat2000,Granger2011} 
Spores were purified as described,\cite{Nicholson1990} lyophilized, and stored 
dry.  
All spores used were free ($> 98$~\%) from growing or sporulating cells, 
germinated spores and cell debris as determined by microscopy.

To prepare NMR samples, the lyophilized spores were suspended in 5~\mM\ sodium 
phosphate buffer in 99.9~\% \ce{D2O} (pH$^\ast$~7.6, where pH$^\ast$ reports the 
reading of the pH meter calibrated with \ce{H2O} buffers) and rehydrated during 
20~h at 6~$\degc$. 
It has been reported that lyophilization and rehydration of spores do not induce 
germination.\cite{Fairhead1994}
The samples were then centrifuged at 5000~$\times$~g for 20~min. 
The supernatant was discarded and the spore pellet was transferred to an 8~mm 
o.d.\ $\times$ 20~mm height NMR tube insert (for relaxation measurements) or to 
a 5~mm o.d.\ $\times$ 20~mm height NMR tube (for quadrupolar echo) and sealed 
with parafilm. 
The insert was placed in a 10~mm o.d.\ NMR tube for relaxation measurements.   

Upon completion of the NMR experiments, the water content of each sample was 
determined gravimetrically by drying at 105~$\degc$ for 16~h. 
The dried spore mass was then used for elemental analysis by inductively coupled 
plasma sector field mass spectrometry (performed at ALS Scandinavia AB, Lule\aa, 
Sweden) and for complete amino acid analysis (performed at Amino Acid Analysis 
Center, Department of Biochemistry and Organic Chemistry, Uppsala University, 
Sweden).
}

%
%
\subsection{\label{subsec:nmr}NMR experiments}

{\small
The longitudinal relaxation of the water \htwo\ magnetization was studied at 
Larmor frequencies from 2.5 to 92.1~MHz using four NMR instruments: 
a Tecmag Discovery spectrometer equipped with a Drusch iron-core magnet 
(2.48 -- 13.1~MHz \htwo\ frequency) or with a Bruker 4.7~T cryomagnet (30.7~MHz) 
and Varian DirectDrive 500 (76.7~MHz) and 600 (92.1~MHz) spectrometers. 
Standard inversion recovery pulse sequences were used. 
All relaxation experiments were performed at $27.0 \pm 0.1~\degc$, maintained by 
a thermostated air-flow. 
The temperature was checked with a thermocouple referenced to an ice-water bath. 
All relaxation experiments were completed within 72~h of sample preparation and 
all samples were examined together at each field. 
Between measurements, the samples were stored at 4~$\degc$. 
No sign of spore germination was observed during the relaxation experiments. 

The \htwo\ quadrupolar echo experiment was performed at 27~$\degc$ on a Bruker 
Avance II 500 spectrometer equipped with a wideline probe (Bruker PH HP BB 500 
SB). 
The quadrupolar echo sequence (90$\degree$ -- $\tau$ -- 90$\degree$ -- $\tau$ -- 
acquisition) was used with 90$\degree$ pulse length of 2.55~\us, echo delay 
$\tau$ of 25~\us, and recycle delay of 120~s. 
The reported spectrum is an average of 2000 transients, accumulated during 67~h 
immediately following sample preparation.  
No sign of spore germination was observed after the NMR experiment.
}

%
%
\section{\label{sec:result}Results and discussion}

%
%
\subsection{\label{subsec:relax}\htwo\ relaxation of core and non-core water}

Water \htwo\ relaxation measurements in the frequency range 2.5 -- 92.1~MHz were 
performed on three \textit{B.\ subtilis} spore preparations that had been 
sporulated at 23, 37, or 43~$\degc$ and on one preparation sporulated in a 
Mn-depleted medium at 37~$\degc$. 
Henceforth, we refer to the former three preparations as native spores, 
labeling the samples N23, N37, and N43, respectively. 
The Mn-depleted spore sample is labeled $-$Mn.

\begin{figure}[!t]
  \centering
  \includegraphics[viewport=0 0 207 205]{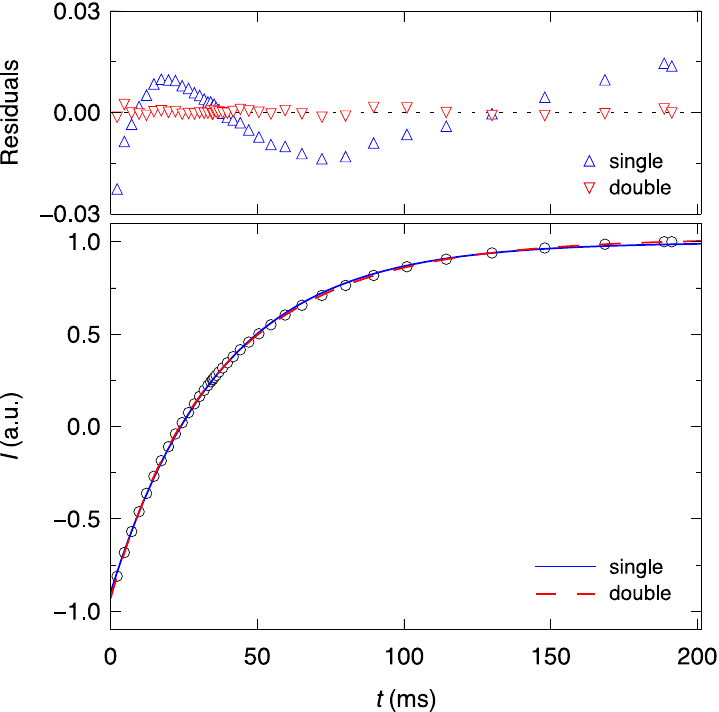}
  \caption{\label{fig:biexp}Water \htwo\ inversion recovery at a magnetic field 
           of 14.1~T (\htwo\ frequency 92.1~MHz) from \textit{B.\ subtilis} spore 
           sample N37. 
           Single-exponential (solid curve) and double-exponential (dashed curve) 
           fits are shown. 
           Note the large and systematic residuals for the single-exponential 
           fit.}
\end{figure}

The water \htwo\ longitudinal magnetization was found to relax bi-exponentially 
at all investigated frequencies (Figs.\ \ref{fig:biexp} and~\ref{fig:indivfit}), 
indicating that the samples contain two water populations in slow (or 
intermediate) exchange on the relaxation time scale.\cite{Zimmerman1957} 
The minor relaxation component (with the smallest weight) has the fastest 
relaxation (the largest longitudinal relaxation rate $R_1$). 
As in the previous MRD study,\cite{Sunde2009b} we assign the minor and major 
components to water in the core (CR) and non-core (NCR) regions, respectively.  
The NCR region comprises all compartments outside the core, including the 
extracellular space. 
This assignment is motivated by the following considerations: (i) the inner 
membrane (IM) is the principal permeability barrier in the 
spore\cite{Setlow2006}; 
(ii) the core is expected to contain less water than the rest of the 
sample\cite{Gerhardt1989,Beaman1986}; 
(iii) the core water is expected to be less mobile than non-core water, 
resulting in a larger $R_1$\cite{Sunde2009b}; 
and (iv) EDTA treatment, which removes \mn\ from the NCR region but not from 
the core, strongly reduces $R_1$ of the major component but has no significant 
effect on $R_1$ of the minor component.\cite{Sunde2009b} 
The observation of slow water exchange across the IM implies that the mean 
residence time of a water molecule in the core is much longer than 10~ms for the 
spore samples examined here.\cite{Sunde2009b,Zimmerman1957}
This result is not inconsistent with the finding, albeit for \textit{Bacillus 
cereus} spores, that \ce{D2O}/\ce{H2O} exchange of the entire spores occurs 
within 1~s.\cite{Kong2013}

\begin{figure}[!t]
  \centering
  \includegraphics[viewport=0 0 201 144]{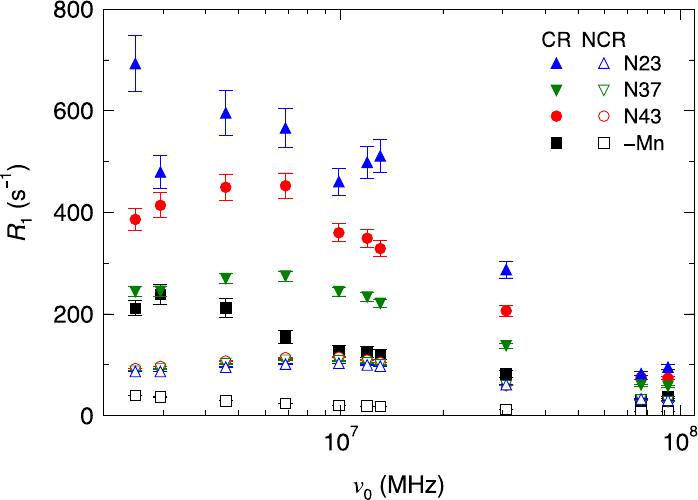}
  \caption{\label{fig:jointfit}\htwo\ MRD profiles for the two water components 
           in native spores sporulated at different temperatures and in Mn-
           depleted spores.  
           The component rates $R_1\rsup{CR}$ and $R_1\rsup{NCR}$ were obtained 
           from global inversion recovery fits, where $f\rsub{CR}$ was 
           constrained to be independent of frequency. 
           The resulting $f\rsub{CR}$ values are given in Table~\ref{tab:sample}. 
           The $R_1\rsup{NCR}$ rate shown here and in Fig.~\ref{fig:pre} has 
           been scaled to $h = 1.5$~g \ce{D2O} (g dry spore mass)$^{-1}$ from 
           the water contents reported in Table~\ref{tab:sample}, assuming that 
           $R_1\rsup{NCR} - R_1^0 \propto 1/h$, where $R_1^0$ is the bulk 
           \ce{D2O} relaxation rate.}
\end{figure}

The bi-exponential form of the inversion recovery is most clearly seen from the 
residuals, which are an order of magnitude larger and show a systematic 
variation in time for the single-exponential fit (Fig.~\ref{fig:biexp}). 
In the slow exchange regime, the relative weight of the minor component, 
$f\rsub{CR}$, can be identified with the water fraction in the core, which must 
be the same at all frequencies.\cite{Zimmerman1957} 
While this is approximately the case, the $f\rsub{CR}$ values obtained from the 
individual fits vary somewhat with frequency (Fig.~\ref{fig:indivfit}b). 
This variation can be attributed to the well-known difficulty of uniquely 
determining the parameters of a bi-exponential decay when the decay rates are 
not very different and one of the components has a small weight ($\sim 0.15$ 
here).  
The ratio, $R_1\rsup{CR}/R_1\rsup{NCR}$, of the component relaxation rates is in 
the range 2 -- 8 (Fig.~\ref{fig:indivfit}a), being smallest at the highest 
frequencies (as in the example shown in Fig.~\ref{fig:biexp}). 
To improve the accuracy of the parameters, we performed global inversion 
recovery fits including data at all frequencies and with $f\rsub{CR}$ 
constrained to be independent of frequency, as expected in the slow exchange 
regime. 
Figure~\ref{fig:jointfit} shows the resulting component MRD profiles for the 
four spore samples, while the $f\rsub{CR}$ values are listed in 
Table~\ref{tab:sample}.

\begin{table}[t]
  \centering
  \begin{threeparttable}
    \caption{\label{tab:sample}Estimates of spore and core water content.}
    \small
    \begin{tabular}{ccccc}
      \toprule
      sample & $h$\,\tnote{a} & $f\rsub{CR}$\tnote{b}
      & $h\,f\rsub{CR}$\tnote{c} & $h\rsub{CR}$\tnote{d} \\
      \midrule
      N23   & 1.59 & $0.122 \pm 0.002$ & $0.194 \pm 0.003$ & 0.70 \\
      N37   & 1.79 & $0.174 \pm 0.008$ & $0.311 \pm 0.014$ & 0.62 \\
      N37   & 1.45 & $0.131 \pm 0.005$ & $0.189 \pm 0.007$ & 0.60 \\
      $-$Mn & 1.39 & $0.139 \pm 0.002$ & $0.169 \pm 0.003$ &      \\
      \bottomrule
    \end{tabular}
    \begin{tablenotes}[flushleft]
      \item[a] Gravimetrically determined sample water content in units of 
               g total \ce{D2O} (g dry spore mass)$^{-1}$.
      \item[b] NMR-derived core water fraction in units of g core \ce{D2O} 
               (g total \ce{D2O})$^{-1}$.
      \item[c] Core water content in units of g core \ce{D2O} (g dry spore 
               mass)$^{-1}$.
      \item[d] Core water content in decoated spores\cite{Melly2002} in 
               units of g core \ce{D2O} (g dry core mass)$^{-1}$.
     \end{tablenotes}
   \normalsize
 \end{threeparttable}
\end{table}

%
%
\subsection{\label{subsec:temp}Sporulation temperature and core water content}

Upon completion of the $R_1$ experiments, the water content $h$ in units of g 
\ce{D2O} (g dry spore mass)$^{-1}$ was determined gravimetrically for each 
sample (Table~\ref{tab:sample}). 
Both $h$ and $f\rsub{CR}$, determined from inversion recovery fits, depend on 
the amount of extracellular water, but the product $h\,f\rsub{CR}$, also given 
in Table~\ref{tab:sample}, does not since it represents the amount of core water 
on a spore dry mass basis.  
In previous work,\cite{Gerhardt1989,Melly2002} the core water content 
$h\rsub{CR}$ on a core dry mass basis, in units of g core \ce{H2O} (g dry core 
mass)$^{-1}$, was determined by density gradient centrifugation after the coat 
and cortex were rendered permeable to the gradient material by chemical 
treatment. 
These values,\cite{Melly2002} interpolated to our temperatures and scaled by a 
factor of 1.111 to convert from \ce{H2O} to {D2O}, are included in 
Table~\ref{tab:sample}. 
The $h\rsub{CR}$ values are higher than the $h\,f\rsub{CR}$ values, as expected 
since the core only accounts for a fraction of the spore's dry mass. 
However, whereas $h\rsub{CR}$ decreases with increasing sporulation temperature, 
we find that $h\,f\rsub{CR}$ varies non-monotonically with sporulation 
temperature, with a $\sim 50$~\% higher value at 37~$\degc$ than at 23 or 
43~$\degc$.

In principle, these two sets of results could be reconciled if the 
(dry core)/(dry spore) mass fraction varies non-monotonically with sporulation 
temperature, being $\sim 0.50$ at 37~$\degc$ and $\sim 0.30$ at the other two 
temperatures. 
Increased sporulation temperature has been shown to reduce the amount of coat 
protein, but has little effect ($\leq 10$~\%) on the levels of DPA and small 
acid-soluble proteins in the core or peptidoglycan in the 
cortex.\cite{Melly2002} 
Another study indicated comparable mineral content in \textit{B.\ subtilis} 
spores sporulated at 25, 30, and 37~$\degc$.\cite{Atrih2001} 
While the core/spore dry mass ratio may thus increase somewhat with increasing 
sporulation temperature, a strong non-monotonic variation seems unlikely. 
Moreover, the amino acid analysis of our samples does not indicate any 
significant differences among the samples in either total protein content 
(Table~\ref{tab:aaa}) or amino acid composition (Table~\ref{tab:protein}), as 
might have been expected if the core/spore dry mass ratio varied by 50~\%.

If the non-monotonic variation of $h\,f\rsub{CR}$ really reflects a 
non-monotonic variation of the core water content (as measured by $h\rsub{CR}$), 
then $R_1\rsup{CR}$ should also vary non-monotonically, with the smallest value 
at the sporulation temperature (37~$\degc$) with the largest core water content 
(where the slowing down of water rotation should be least pronounced). 
This is indeed observed (Fig.~\ref{fig:jointfit}). 
Some caution is required here, however, since $R_1\rsup{CR}$ is the sum of a 
quadrupolar contribution, which reflects the average water mobility in the core, 
and a paramagnetic relaxation enhancement (PRE) contribution (see below). 
Both of these contributions are expected to decrease with increasing core water 
content, but the PRE contribution also depends on the Mn content of the core. 
According to the elemental analysis (Table~\ref{tab:element}), the Mn content 
does not differ significantly between samples N23 and N37, whereas sample N43 
seems to have a slightly lower Mn content, which might explain the difference in 
$R_1\rsup{CR}$ between samples N23 and N43 (Fig.~\ref{fig:jointfit}).

In conclusion, this analysis suggests that the water content of the core 
(relative to the dry mass of the core) is indeed significantly higher in sample 
N37 than in samples N23 and N43. 
However, a non-monotonic variation of core water content with sporulation 
temperature is \textit{a priori} unlikely and contradicts previous results from 
decoated spores.\cite{Gerhardt1989,Melly2002} 
At this stage, we would therefore not rule out the possibility that other 
factors, besides the sporulation temperature, affected the core water content of 
our samples. 
In this regard, we note that sample N37 deviates somewhat from the other two 
samples in the elemental analysis (notably for Si and Fe, see 
Table~\ref{tab:element}), although the significance of these differences is 
uncertain. 
We also note that any correlation with sporulation temperature in our study 
hinges on the assumption that the core water content at sporulation is 
reproduced after rehydration of the freeze-dried spores.

%
%
\subsection{\label{subsec:mn}Hydration of \mn\ in the core}

In the previous MRD study of \textit{B.\ subtilis} spores, manganese was partly 
removed from the spores by EDTA treatment.\cite{Sunde2009b} 
This treatment only reduced the total Mn content of the spore by 32~\% and it 
was concluded that the \mn\ ions in the core were essentially 
unaffected.\cite{Sunde2009b} 
In contrast, in the –Mn sample investigated here, the total Mn content is 
reduced by a factor of $181 \pm 24$ (Table~\ref{tab:element}). 
More than 99~\% of the core \mn\ is thus removed. 
In the $-$Mn sample, we expect that core \mn\ is replaced by \ca\ and this 
expectation is supported by the elemental analysis (Table~\ref{tab:element}). 
It has been shown previously that these Mn-depleted spores have normal core 
water and DPA contents and normal heat resistance.\cite{Granger2011}

For the three native spore samples, the water \htwo\ MRD profiles in 
Fig.~\ref{fig:jointfit} exhibit maxima just below 10~MHz in both the NCR and CR 
components. 
These maxima indicate a Mn(II)-induced PRE contribution to 
$R_1$.\cite{Koenig1985,Bertini2005} 
In contrast, the MRD profiles from the $-$Mn spores do not show such a maximum 
(Fig.~\ref{fig:jointfit}), demonstrating that \mn\ is the dominant source of the 
PRE in the native spores. 
In the previous MRD study,\cite{Sunde2009b} inversion recovery experiments were 
only performed down to 12~MHz so the PRE maximum could only be observed in the 
effective $R_1$ measured by the field-cycling technique, which essentially 
reports on the dominant NCR component. 
Based on the $R_1\rsup{CR}$ data above 12~MHz, it was concluded that there is 
little or no direct \mn--water contact in the core.\cite{Sunde2009b} 
The maximum observed in the more extensive $R_1\rsup{CR}$ data reported here 
prompt us to revise this conclusion.

\begin{figure}[t]
  \includegraphics[viewport=0 0 201 144]{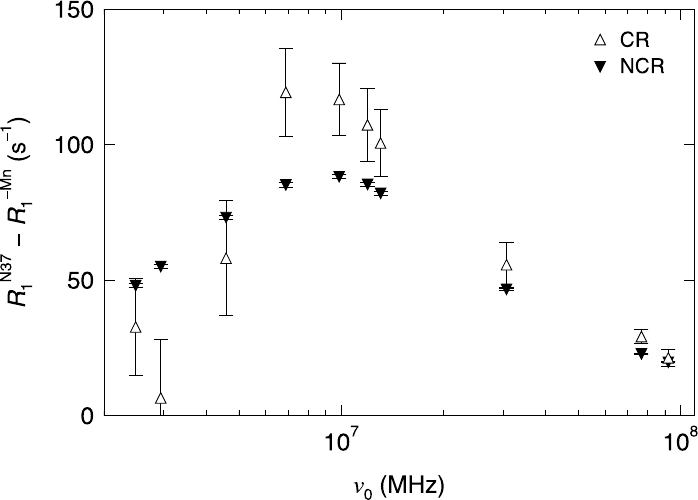}
  \caption{\label{fig:pre}PRE contribution to \htwo\ relaxation in the core and 
           NCR regions, isolated by taking the difference of the respective 
           component rates in Fig.~\ref{fig:jointfit} for the N37 and 
           $-$Mn samples.}
\end{figure}

To highlight the PRE contribution, we plot in Fig.~\ref{fig:pre} the differences 
$R_1\rsup{CR}(\mathrm{N37}) - R_1\rsup{CR}(-\mathrm{Mn})$ and 
$R_1\rsup{NCR}(\mathrm{N37}) - R_1\rsup{NCR}(-\mathrm{Mn})$. 
Even though the \mn\ concentration is much higher in the core than in the NCR 
region, the PRE contributions are similar in the two regions 
(Fig.~\ref{fig:pre}). 
The data in Fig.~\ref{fig:pre} yield $1.04 \pm 0.24$ for the CR/NCR ratio of PRE 
contributions. 
This observation indicates that most \mn\ ions in the core are not in direct 
contact with water or else that the coordinating water molecules exchange so 
slowly ($\gg 10$~\us) that they do not contribute to the PRE. 
For a quantitative estimate, we note that the PRE contribution to $R_1$ is 
proportional to $q\rsub{Mn}/N\rsub{Mn}$, where $q\rsub{Mn}$ is the effective 
water coordination number of the \mn\ ions and $N\rsub{Mn}$ is the water/Mn mole 
ratio. 
Furthermore, $N\rsub{Mn}(\mathrm{NCR})/N\rsub{Mn}(\mathrm{CR}) 
= [p\rsub{CR}/(1 - p\rsub{CR})] \times [(1 - f\rsub{CR})/f\rsub{CR}] 
\approx (0.68/0.32) \times (0.85/0.15) \approx 12$. 
Here, we have assumed that the Mn fraction in the core is $p\rsub{CR} = 0.68$, 
as determined previously,\cite{Sunde2009b} and we have used the average water 
fraction $f\rsub{CR}$ for samples N37 and $-$Mn (Table~\ref{tab:sample}). 
Given that $q\rsub{Mn}(\mathrm{NCR}) = 1.1 \pm 0.2$, as determined 
previously using the conventional Solomon--Bloembergen--Morgan 
theory,\cite{Sunde2009b} we thus arrive at $q\rsub{Mn}(\mathrm{CR}) 
= q\rsub{Mn}(\mathrm{NCR}) \times 1.04 / 12 = 0.10 \pm 0.03$. 
In other words, the direct water coordination of \mn\ ions in the core is 
only 1.7~\% of that for fully hydrated ions (with $q\rsub{Mn} = 6)$. 
This conclusion is consistent with the observation that the manganese EPR 
spectrum from \textit{Bacillus megaterium} spores closely resembles that from an 
anhydrous 10:1 Ca:Mn DPA chelate model system\cite{Johnstone1982} and with \ctt\ 
NMR spectra indicating that the DPA in the core is in a solid-like 
state.\cite{Leuschner2000}    

%
%
\subsection{\label{subsec:mobility}Water mobility in the core}

The previous MRD study of \textit{B.\ subtilis} spores concluded that the 
(rotational) mobility of water in the core is slowed down by a factor of 
$30.8 \pm 0.5$ on average compared to bulk water.\cite{Sunde2009b} 
As noted,\cite{Sunde2009b} this estimate is an upper bound, since it was 
obtained by neglecting any PRE contribution to $R_1\rsup{CR}$. 
Indeed, the present results demonstrate a significant PRE contribution in the 
core. 
A more accurate estimate of core water mobility can now be obtained from 
$R_1\rsup{CR}$ for the $-$Mn sample, which does not have a PRE contribution. 
Dividing the high-frequency $R_1\rsup{CR}$ (averaging over the two highest 
frequencies in Fig.~\ref{fig:biexp}a) by the bulk \ce{D2O} relaxation rate at 
27~$\degc$ ($R_1^0 = 2.20~\pers$), we obtain a slowing-down factor of $15.3 \pm 
3$. 
This result happens to coincide with the slowing-down factor of $15.6 \pm 3$ 
reported for the macromolecular hydration layers of an \textit{Escherichia coli} 
cell.\cite{Persson2008b} 
Because the rotational correlation of bulk \ce{H2O} is 1.6~ps at 27~$\degc$, 
this corresponds to an average rotational correlation time for core water of 
25~ps. 
Core water is thus highly mobile.  

Although the detailed structure of the core is unknown, its high content of 
proteins, nucleic acids, DPA, and calcium ions implies a pronounced 
microheterogeneity and a correspondingly wide distribution of water mobility. 
The slowing-down factor of 15 is the mean of this wide distribution. 
It is therefore dominated by the slowest water molecules up to $\sim 2$~ns, 
which is the cutoff set by the highest examined resonance 
frequency.\cite{Sunde2009b} 
Consequently, the mobility of most of the core water is likely to be much closer 
to the bulk-water value than suggested by the average slowing-down factor.       

For the NCR region, the high-frequency $R_1\rsup{NCR}$ yields a slowing-down 
factor of 3.6 and, after correcting for the extracellular 
water,\cite{Sunde2009b,Nakashio1985} a slowing-down factor of $5 \pm 1$ for the 
cortex and coat regions of the spore. 
Essentially the same result\cite{Eijlander2011,Driks1999} was obtained 
previously based on $R_1\rsup{NCR}$ from EDTA-treated spores.\cite{Sunde2009b} 
The agreement is consistent with the expectation that the NCR region of 
EDTA-treated spores has the same low Mn content as in the Mn-depleted spores 
studied here, so that the PRE contribution to $R_1\rsup{NCR}$ is negligibly 
small in both cases. 
The slowing-down factor of five for the cortex and coat regions of the spore is 
comparable to the factor of three deduced in a similar way for the cell water of 
\textit{E.\ coli}.\cite{Persson2008b} 
Being the mean of wide distribution, the factor of five is presumably strongly 
influenced by the minor fraction of water confined to the dense coat region. 
The more water-rich cortex should thus have a water slowing-down factor 
significantly smaller than five, probably five- to ten-fold smaller than the 
slowing-down factor of 15 for the core. 
In contrast, the spin-probe EPR study mentioned in Sec.~\ref{sec:intro} inferred 
only a two-fold mobility difference between core and cortex.\cite{deVries2006}

If, as previously argued,\cite{Gould1986,Sapru1993,Ablett1999,Rice2011} the core 
were in a glassy state, then the core water would presumably be in the `rigid 
lattice' NMR regime (correlation time $\gg 10^{-6}$~s), producing a 100 -- 
200~kHz wide \htwo\ NMR spectrum that would not contribute to the \htwo\ NMR 
signal detected in relaxation experiments. 
Therefore, although our \htwo\ relaxation data demonstrate that at least a large 
fraction of the core water is highly mobile, they do not exclude the possibility 
that a small fraction of core water is immobilized.

\begin{figure}[t]
  \includegraphics[viewport=0 0 199 207]{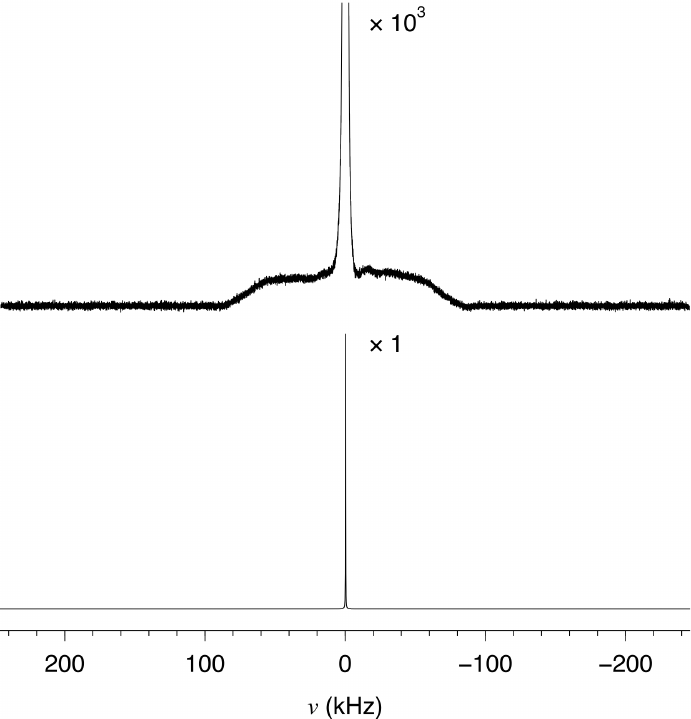}
  \caption{\label{fig:qecho}\htwo\ quadrupolar echo spectrum from a Mn-depleted 
           spore sample with $h = 1.25$~g \ce{D2O} (g dry spore mass)$^{-1}$). 
           To reveal the broad component, the full spectrum (bottom) was 
           vertically magnified by a factor of 1000 while truncating the central 
           peak (top).}
\end{figure}

To detect any immobilized water in the core (or elsewhere in the spore), we 
recorded a \htwo\ quadrupolar echo spectrum from a Mn-depleted spore sample 
(Fig.~\ref{fig:qecho}). 
The spectrum is a superposition of a narrow central peak and a much weaker broad 
component. 
The narrow peak originates from the mobile core and NCR water that we have 
characterized by MRD measurements. 
The shape and width of the broad component is consistent with exchangeable ND 
and OD deuterons (with a distribution of orientational order parameters) in 
immobilized macromolecules,\cite{Usha1989,Usha1991,Venu1999} henceforth referred 
to as labile deuterons (LDs). 
But the broad component is also consistent with immobilized \ce{D2O} molecules 
or with a combination of LDs and immobilized water. 
It is not possible to discriminate among these possibilities solely on the basis 
of the spectral lineshape. 
However, as we shall show, the relative intensity of the broad component 
indicates that it is produced by LDs rather than by immobilized water.

To estimate the relative intensities of the two spectral components, 
proportional to the number of contributing deuterons, we (separately) fitted a 
Lorentzian lineshape to the narrow peak and a spin-1 powder pattern (a so-called 
Pake doublet) to the broad spectral component. 
Due to the finite pulse length (see Sec.~\ref{subsec:nmr}), the broad component 
is not fully excited and its intensity is therefore reduced by a factor of 
0.76.\cite{Usha1989,Usha1991} 
Taking this factor into account, we find that the broad component corresponds to 
$6.7 \pm 0.3$~\% of the deuterons in the sample. 
The water content of the Mn-depleted sample used for the quadrupolar echo 
experiment was $h = 1.25$~g \ce{D2O} (g dry spore mass)$^{-1}$. 
Using also the results of the amino acid analysis (Tables \ref{tab:aaa} 
and~\ref{tab:protein}), we estimate that 5.8~\% of the deuterons in the sample 
are ND or OD deuterons in spore proteins. 
The proteins are largely confined to the core and coat regions of the spore, 
where they are expected to be rotationally immobilized.\cite{Sunde2009b,
Henriques2000} 
The 53~\% non-protein part of the dry spore mass (Table~\ref{tab:protein}) 
includes (immobilized) DNA in the core and (partly immobilized) peptidoglycan in 
the cortex as well as components, such as DPA and calcium in the core, that do 
not contain LDs. 
The relative intensity of the broad component can thus be fully accounted for by 
LDs in the immobilized protein, nucleic acid, and peptidoglycan in the spore.

It is conceivable that a small fraction of the core water is immobilized, but 
the presence or absence of such a small amount of immobilized water cannot be 
demonstrated from \htwo\ spectral analysis, since the LD contribution is not 
accurately known. 
For example, if 10~\% of the core water were immobilized, this would merely 
increase the relative intensity of the broad spectral component by $\sim 
1.5$~\%, which is comparable to the uncertainty in the LD contribution. 
In other words, the \htwo\ spectrum in Fig.~\ref{fig:qecho} establishes an upper 
bound of $\sim 10$~\% for the fraction of immobilized water in the core.

In contrast to our analysis, a \htwo\ quadrupolar echo spectrum from 
\textit{B.\ subtilis} spores reported by Rice \textit{et al.}\ was taken as 
evidence for a significant fraction of immobilized spore water.\cite{Rice2011} 
Whereas our spore samples are fully hydrated, the sample studied by these 
authors was vacuum (0.05~mbar) dried for 24~h. 
The water content of this sample was not reported, but the authors claimed that 
they only removed ``excess water''.\cite{Rice2011} 
The relative intensities of the spectral components were not determined, but the 
reported spectrum suggests that the narrow central peak, due to mobile water, 
accounts for at most a few percent of the deuterons in the 
sample.\cite{Rice2011} 
In contrast, the central peak accounts for $93.3 \pm 0.3$~\% of the deuterons in 
our sample, which includes 25 -- 30~\% extracellular 
water.\cite{Sunde2009b,Nakashio1985} 
It is clear, therefore, that the drying protocol used by Rice \textit{et al.}\ 
removes not only extracellular water but also most of the water inside the spore. 
Such extensive dehydration may produce structural changes that immobilize a 
significant fraction of the residual water. 
However, we do not believe that the spectrum reported by Rice \textit{et al.}\ 
can be taken as evidence for immobilized water even in their dehydrated spores.

The \htwo\ spectrum from dehydrated \textit{B.\ subtilis} spores features, in 
addition to the small central peak, a major broad component with a resolved 
quadrupole splitting of 130~kHz and a minor broad component with a resolved 
quadrupole splitting of 36~kHz.\cite{Rice2011} 
Rice \textit{et al.}\ assigned the major 130~kHz component to immobilized water 
and the minor 36~kHz component to LDs. 
Since the water content of the dehydrated sample was not determined, it is not 
possible to base the assignment on the relative intensity. 
Without further information, it is, in our view, more reasonable to assign the 
major 130~kHz component to LDs (which must contribute) rather than to 
immobilized water (which is conjectural). 
For the origin of the minor 36~kHz component, we propose LDs in the (partly 
flexible) cortex peptidoglycan and/or in \ce{CD3} and \ce{ND3} groups, the fast 
internal rotation of which reduces the \htwo\ quadrupole splitting by a factor 
of three. 
(The spores studied by Rice \textit{et al.}\ were prepared from bacteria 
cultivated in a partly deuterated medium so that deuterons were also 
biosynthetically incorporated in nonlabile macromolecular positions.) 
In conclusion, we do not believe that the spectrum reported by Rice 
\textit{et al.}\ demonstrates the existence of immobilized water in either 
dehydrated or fully hydrated spores.

%
%
\section{\label{sec:conc}Conclusions}

In this work, we have characterized the state of water in the core of native and 
Mn-depleted \textit{B.\ subtilis} spores with the aid of two well-established 
\htwo\ NMR experiments. 

Multiple-field inversion recovery experiments were performed for the first time 
in the frequency range where the paramagnetic enhancement of water \htwo\ 
relaxation due to \mn\ ions in the core can be identified. 
This enhancement allowed us to quantify the \mn–water contact in the core, 
previously thought to be negligibly small.\cite{Sunde2009b} 
In addition, by examining spores with $> 99$~\% of the core \mn\ substituted by 
\ca, we could measure the \htwo\ relaxation rate of core water free from 
paramagnetic effects. 
The characterization of core water mobility in the previous MRD 
study\cite{Sunde2009b} could thereby be substantially improved.

A \htwo\ quadrupolar echo experiment was performed for the first time on fully 
hydrated spores and we showed that the resulting solid-state \htwo\ NMR spectrum 
can be accounted for by labile macromolecular deuterons, without invoking 
immobilized water in the core or elsewhere in the spore. 
This result differs qualitatively from the conclusion of a recent solid-state 
\htwo\ NMR study of dehydrated spores.\cite{Rice2011}

We also examined the effect of sporulation temperature, but we could not confirm 
the expected reduction of core water content with increasing sporulation 
temperature.\cite{Melly2002} 
The origin of this discrepancy is unclear, but we speculate that other 
parameters, besides sporulation temperature, affect the core water content in 
our samples.

The three most significant quantitative findings of this work are as follows.
\begin{enumerate}[label=(\roman*),series=conc]
  \item Fully hydrated spores do not contain detectable amounts of immobilized 
        water. 
        This means that any immobilized water amounts to less than $\sim 1$~\% 
        of the spore water or less than $\sim 10$~\% of the core water.     
  \item The rotational motion of the observable mobile core water is slowed down 
        by a factor of 15 as compared to bulk water. 
        This average value is likely dominated by a small fraction of core water 
        with rotational correlation time of order 1~ns, while water rotation 
        occurs on a time scale of less than 10~ps in the vast majority of 
        hydration sites in the core.
  \item The large depot of manganese in the core is nearly anhydrous, the direct 
        water contact amounting, on average, to 1.7~\% of the sixfold 
        coordination of fully solvated \mn.
\end{enumerate}

In conclusion, the \htwo\ NMR data presented here fully support the gel scenario 
for the core of \textit{B.\ subtilis} spores, and are clearly inconsistent with 
the glass scenario.

%
%
\section*{Acknowledgments}

{\small
We thank Sanna Gustavsson and Daniel Topgaard for help with the solid-state NMR 
measurements. 
This work was supported by the Nestl\'e Research Center (B.H.\ and S.K.), 
the Knut and Alice Wallenberg Foundation (B.H.), 
and a Department of Defense Multi-disciplinary University Research Initiative 
through the U.S.\ Army Research Laboratory and the U.S.\ Army Research Laboratory 
under contract number W911NF-09-1-0286 (B.S.\ and P.S.).  
}

%
%

%
%

\setcounter{figure}{0}
\setcounter{table}{0}

\makeatletter
\renewcommand{\thefigure}{S\arabic{figure}}
\renewcommand{\thetable}{S\arabic{table}}
\makeatother

\onecolumn
\section*{Supporting material}

\vspace*{\fill}
\begin{figure}[!h]
  \centering
  \includegraphics[viewport=0 0 211 272]{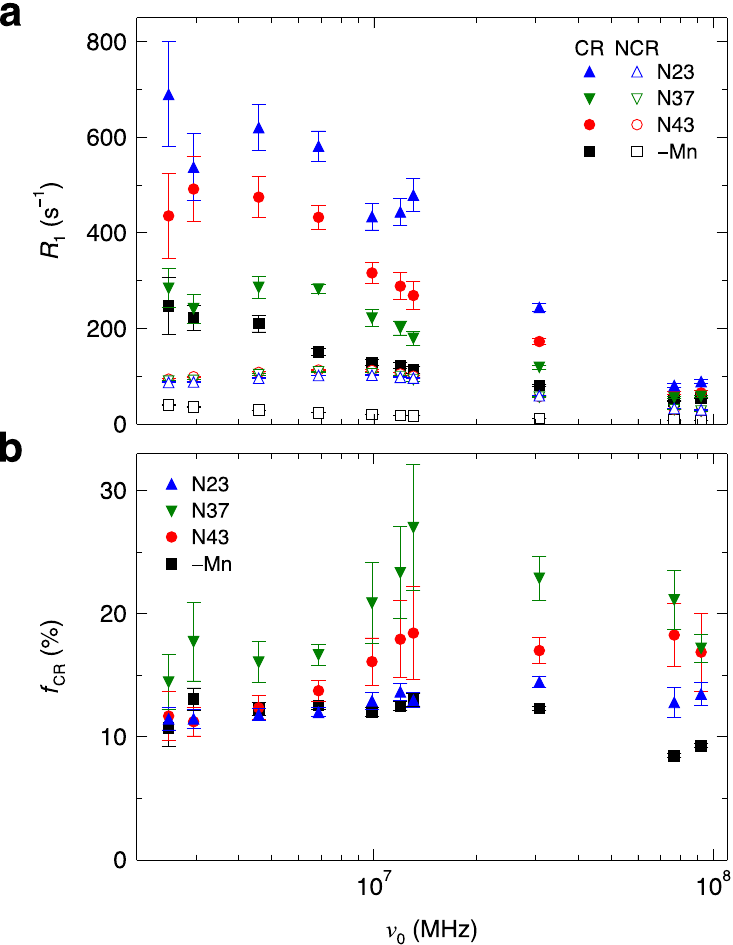}
  \caption{\label{fig:indivfit}Results of individual bi-exponential fits to the 
           water \htwo\ inversion recovery from native spores with different 
           sporulation temperatures and from Mn-depleted spores: 
           (\textbf{a}) component relaxation rates, $R_1\rsup{CR}$ and 
           $R_1\rsup{NCR}$ and (\textbf{b}) the relative weight, $f\rsub{CR}$, 
           of the minor component. 
           The inversion recovery data sets at each frequency were fitted 
           separately, without constraining $f\rsub{CR}$ to be independent of 
           frequency. 
           The $R_1\rsup{NCR}$ rate in panel (\textbf{a}) has been scaled to 
           $h = 1.5$~g \ce{D2O} (g dry spore mass)$^{-1}$ from the water contents 
           reported in Table~\ref{tab:sample}, assuming that 
           $R_1\rsup{NCR} - R_1^0 \propto 1/h$, where $R_1^0$ is the bulk 
           \ce{D2O} relaxation rate.}
\end{figure}

\vspace*{\fill}

\newpage

\vspace*{\fill}

\begin{table}[!h]
  \centering
  \begin{threeparttable}
    \caption{\label{tab:element}Elemental abundance 
             ({\tmu}mol (g dry mass)$^{-1}$) in spore samples.$\rsup{a}$}
    \small
    \begin{tabular}{lcccc}
      \toprule
      & N23 & N37 & N43 & $-$Mn \\
      \midrule
      Ca & 760             & 800             & 830             & 960 \\
      Fe & 0.64            & 0.20            & 0.60            & 1.0 \\
      K  & 100             & 78              & 104             & 95 \\
      Mg & 87              & 104             & 88              & 148 \\
      Mn & $189 \pm 17$    & $184 \pm 17$    & $165 \pm 15$    & $0.91 \pm 0.08$\\
      Si & 11              & 1.8             & 21              & 17 \\
      Zn & $0.61 \pm 0.06$ & $0.64 \pm 0.06$ & $0.80 \pm 0.08$ & $0.65 \pm 0.06$\\
      \bottomrule
    \end{tabular}
    \begin{tablenotes}[flushleft]
      \footnotesize
      \item[a] When given, uncertainties represent one standard deviation. 
               When not given, the analysis is less accurate.
    \end{tablenotes}
  \normalsize
  \end{threeparttable}
\end{table}

\vspace*{\fill}

\begin{table}[!h]
  \centering
  \begin{threeparttable}
    \caption{\label{tab:aaa}Amino acid composition (mole \%) of spore samples.}
    \small
    \begin{tabular}{lcccc}
      \toprule
      & N23 & N37 & N43 & $-$Mn \\
      \midrule
      Ala                 & 12.0  & 11.4  & 11.0  & 11.3  \\
      Arg                 & 5.1   & 5.2   & 5.1   & 5.2   \\
      Asx\tnote{a} & 8.9   & 9.0   & 9.0   & 8.5   \\
      Cys\tnote{b} & (0.8) & (0.8) & (0.8) & (0.8) \\
      Glx\tnote{c} & 13.2  & 12.9  & 12.5  & 12.6  \\
      Gly                 & 11.0  & 11.7  & 11.8  & 11.2  \\
      His                 & 2.7   & 2.9   & 2.8   & 2.8   \\
      Ile                 & 3.7   & 3.7   & 3.6   & 3.6   \\
      Leu                 & 5.3   & 5.2   & 5.2   & 5.0   \\
      Lys                 & 6.0   & 6.1   & 6.6   & 6.4   \\
      Met                 & 1.5   & 0.5   & 0.4   & 1.6   \\
      Phe                 & 4.4   & 4.5   & 4.4   & 4.4   \\
      Pro                 & 3.6   & 3.1   & 3.1   & 3.6   \\
      Ser                 & 6.1   & 6.5   & 6.8   & 6.6   \\
      Thr                 & 4.3   & 4.4   & 4.4   & 4.3   \\
      Trp\tnote{b} & (1.0) & (1.0) & (1.0) & (1.0) \\
      Tyr                 & 5.1   & 5.7   & 6.3   & 5.8   \\
      Val                 & 5.4   & 5.3   & 5.4   & 5.2   \\
      \bottomrule
    \end{tabular}
    \begin{tablenotes}[flushleft]
      \item[a] Asx = Asn + Asp.
      \item[b] See Table~S3 in Ref.~\citenum{Sunde2009b}.
      \item[c] Glx = Gln + Glu.
    \end{tablenotes}
  \normalsize
  \end{threeparttable}
\end{table}

\vspace*{\fill}

\begin{table}[!h]
  \centering
  \begin{threeparttable}
    \caption{\label{tab:protein}Protein content of spore samples.}
    \small
    \begin{tabular}{lcccc}
      \toprule
      & N23 & N37 & N43 & $-$Mn \\
      \midrule
      protein content & \multirow{2}{*}{0.457} & \multirow{2}{*}{0.457} 
      & \multirow{2}{*}{0.453} & \multirow{2}{*}{0.469} \\
      (g\:(g dry mass)$^{-1}$\,\tnote{a} & & & &  \\
      mass-weighted residue & \multirow{2}{*}{109.5} & \multirow{2}{*}{109.5} 
      & \multirow{2}{*}{109.7} & \multirow{2}{*}{109.9} \\
      molar mass (g\:mol$^{-1}$)\,\tnote{b} & & & & \\
      \bottomrule
    \end{tabular}
    \begin{tablenotes}[flushleft]
      \footnotesize
      \item[a] Including Cys and Tryp with abundance from Table~\ref{tab:aaa}.
      \item[b] Calculated from the data in Table~\ref{tab:aaa}.
    \end{tablenotes}
  \normalsize
  \end{threeparttable}
\end{table}

\vspace*{\fill}

\end{document}